\documentclass[10pt,twocolumn]{IEEEtran}
\usepackage[latin1]{inputenc}
\usepackage{amsmath}
\usepackage{amsbsy}
\usepackage{amssymb}
\usepackage{latexsym}
\usepackage{times}
\usepackage{epsfig}
\title{ Robust Low-Rank LCMV Beamforming Algorithms Based on Joint Iterative Optimization Strategies}

\author{Rodrigo C. de Lamare \\
\thanks{R. C. de Lamare is with the
Communications Research Group, Department of Electronics, University
of York, York Y010 5DD, United Kingdom . E-mail:
rcdl500@ohm.york.ac.uk} }

\begin{document}

\maketitle

\begin{abstract}
This chapter presents \index{robust} reduced-rank linearly
constrained minimum variance (LCMV) \index{beamforming} algorithms
based on the concept of joint iterative optimization of parameters.
The proposed \index{robust} reduced-rank scheme is based on a
constrained robust joint iterative optimization (RJIO) of parameters
according to the minimum variance criterion. The robust optimization
procedure adjusts the parameters of a rank-reduction matrix, a
reduced-rank beamformer and the diagonal loading in an alternating
manner. LCMV expressions are developed for the design of the
rank-reduction matrix and the reduced-rank beamformer. Stochastic
gradient and recursive least-squares adaptive algorithms are then
devised for an efficient implementation of the RJIO robust
beamforming technique. Simulations for a \index{beamforming}
application in the presence of uncertainties show that the RJIO
scheme and algorithms outperform in convergence and tracking
performances existing algorithms while requiring a comparable
complexity.

\begin{keywords}
Adaptive beamforming, constrained optimization, robust techniques,
reduced-rank methods, iterative methods.
\end{keywords}
\end{abstract}

\section{Introduction}

In the last decade, \index{adaptive beamforming} techniques have
attracted a significant interest from researchers and engineers, and
found applications in radar, sonar, wireless communications and
seismology \cite{vantrees,li}. The optimal linearly constrained
minimum variance (LCMV) beamformer is designed in such a way that it
minimizes the array output power while maintaining a constant
response in the direction of a signal of interest (SoI)
\cite{vantrees,li,haykin}. However, this technique requires the
computation of the inverse of the input data covariance matrix and
the knowledge of the array steering vector. Adaptive versions of the
LCMV beamformer were subsequently reported with stochastic gradient
(SG) \cite{frost72,delamaretsp} and recursive least squares (RLS)
\cite{romano96} algorithms. A key problem with \index{adaptive
beamforming} techniques is the impact of uncertainties which can
result in a considerable performance degradation. These mismatches
are caused by local scattering, imperfectly calibrated arrays,
insufficient training and imprecisely known wave field propagation
conditions \cite{li}.

In the last decades a number of robust approaches have been reported
that address this problem \cite{cox}-\cite{fa10}. These techniques
can be classified according to the approach adopted to deal with the
mismatches: techniques based on diagonal loading
\cite{cox,vorobyov03,lorenz,chen07}, methods that estimate the
mismatch or equivalently the actual steering vector
\cite{li03,stoica04,hassanien}, and techniques that exploit
properties such as the constant modulus of the signals
\cite{lei09,lei10,barc} and the low-rank of the interference
subspace \cite{feldman},\cite{scharf}-\cite{fa10}. Furthermore,
beamforming algorithms usually have a trade-off between performance
and computational complexity which depends on the designer's choice
of the adaptation algorithm \cite{haykin,veen}. A number of robust
designs can be cast as optimization problems which end up in the
so-called second-order cone (SOC) program, which can be solved with
interior point methods and have a computational cost that is super
cubic in the number of parameters of the beamformer. This poses a
problem for beamforming systems that have a large number of
parameter and operate in time-varying scenarios, which require the
beamformer to be recomputed periodically.

A robust technique for short-data record scenarios is reduced-rank
signal processing \cite{scharf}-\cite{fa10}, which is very well
suited for systems with a large number of parameters. These
algorithms are robust against short data records, have the ability
to exploit the low-rank nature of the signals encountered in
beamforming applications and can resist moderate steering vector
mismatches. These methods include the computationally expensive
eigen-decomposition techniques \cite{scharf}-\cite{bar-ness} to
alternative approaches such as the auxiliary-vector filter (AVF)
\cite{pados99},\cite{qian}, the multistage Wiener filter (MSWF)
\cite{reed98}, \cite{goldstein}, \cite{santos} which are based on
the Krylov subspace, and joint iterative optimization (JIO)
approaches \cite{hua,delamarespl07,delamareelb,delamare10,jidf}. The
JIO techniques reported in
\cite{delamarespl07,delamareelb,delamare10} outperform the
eigen-decomposition- and Krylov-based methods and are amenable to
efficient adaptive implementations. However, robust versions of JIO
methods have not been considered so far.

In this chapter, robust LCMV reduced-rank beamforming algorithms
based on constrained robust joint iterative optimization (RJIO) of
parameters are developed. The basic idea of the RJIO approach is to
design a bank of robust adaptive beamformers which is responsible
for performing dimensionality reduction, whereas the robust
reduced-rank beamformer effectively forms the beam in the direction
of the SoI and takes into account the uncertainty. Robust LCMV
expressions for the design of the rank reduction matrix and the
reduced-rank beamformer are proposed that can appropriately deal
with array steering vector mismatches. SG and RLS algorithms for
efficiently implementing the method are then devised. An automatic
rank adaptation algorithm for determining the most adequate rank for
the RJIO algorithms is described. A simulation study of the proposed
RJIO algorithms and existing techniques is considered.

This chapter is organized as follows. The system and signals models
are described in Section II. The full-rank and the reduced-rank LCMV
filtering problems are formulated in Section III. Section IV is
dedicated to the RJIO method, whereas Section V is devoted to the
derivation of the adaptive SG and RLS algorithms, the analysis of
the computational complexity, and the rank adaptation technique.
Section VI presents and discusses the simulation results and Section
VII gives the concluding remarks.

\section{System Model}

Let us consider a sensor-array system equipped with a uniform linear
array (ULA) of $M$ elements, as shown in Fig. 1. Assuming that the
sources are in the far field of the array, the signals of $K$
narrowband sources impinge on the array $\left( K < M \right)$ with
unknown directions of arrival (DOA) ${\theta}_l$ for $l=1,2, \ldots,
K$.

\begin{figure}[!htb]
\begin{center}
\def\epsfsize#1#2{1\columnwidth}
\epsfbox{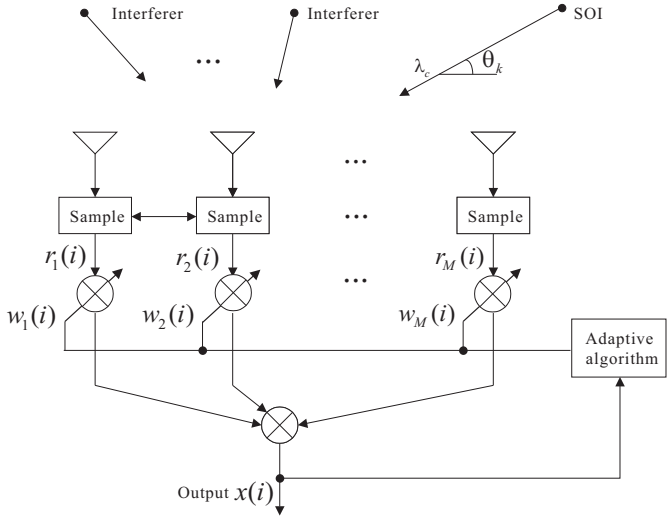} \vspace*{-1em}\caption{Block diagram of a
sensor-array array system with interfering signals}
\end{center}
\end{figure}

The input data from the antenna array can be organized in an $M
\times 1$ vector expressed by
\begin{equation}
{\boldsymbol r}(i) = {\boldsymbol A}(\theta) {\boldsymbol s}(i) +
{\boldsymbol n}(i)
\end{equation}
where $${\boldsymbol A}(\theta) = \left[ {\boldsymbol
a}(\theta_1), \ldots, {\boldsymbol a}(\theta_K)\right] $$ is the
$M \times K$ matrix of signal steering vectors.  The $M \times 1$
signal steering vector is defined as
\begin{equation}
{\boldsymbol a}(\theta_l) = \left[ 1, e^{-2\pi j
\frac{d_s}{\lambda_c}\cos \theta_l}, \ldots, e^{-2\pi j
(M-1)\frac{d_s}{\lambda_c}\cos \theta_l}\right] ^T
\end{equation}
for a signal impinging at angle $\theta_l$, $l=1,2, \ldots, K$,
where $d_s = \lambda_c / 2$ is the inter-element spacing,
$\lambda_c$ is the wavelength and $(.)^T$ denotes the transpose
operation. The vector ${\boldsymbol n}(i)$ denotes the complex
vector of sensor noise, which is assumed to be zero-mean and
Gaussian with covariance matrix $\sigma^2 {\boldsymbol I}$.

\section{Problem Statement and Design of Adaptive Beamformers}

In this section, the problem of designing \index{robust beamforming}
algorithms against steering vector mismatches is stated. The design
of robust full-rank and reduced-rank LCMV beamformers is introduced
along with the modelling of steering vector mismatches. The presumed
array steering vector for the k-th desired signal is given by
${\boldsymbol a}_p (\theta_k) ={\boldsymbol a}(\theta_k) +
{\boldsymbol e}$, where ${\boldsymbol e}$ is the $M \times 1$
mismatch vector and ${\boldsymbol a}(\theta_k)$ is the actual array
steering vector which is unknown by the system. By using the
presumed array steering vector ${\boldsymbol a}_p (\theta_k)$, the
performance of a conventional LCMV beamformer can be degraded
significantly. The problem of interest is how to design a beamformer
that can deal with the mismatch and minimize the performance loss
due to the uncertainty.

\subsection{Adaptive LCMV Beamformers}

In order to perform beamforming with a full-rank LCMV beamformer, we
linearly combine the data vector ${\boldsymbol
r}(i)=[r_{1}^{(i)}~r_{2}^{(i)}~ \ldots ~r_{M}^{(i)}]^{T}$ with the
full-rank beamformer ${\boldsymbol w}= [w_1^{}~ w_2^{} ~ \ldots ~
w_M^{}]^T$ to yield
\begin{equation}
x(i) = {\boldsymbol w}^{H} {\boldsymbol r}(i)
\end{equation}

The optimal LCMV beamformer is described by the $M \times 1$ vector
${\boldsymbol w}$, which is designed to solve the following
optimization problem
\begin{equation}
\begin{split}
{\rm minimize} ~  E[|{\boldsymbol w}^H {\bf r}(i) |^2] & = {\boldsymbol w}^{H} {\boldsymbol R} {\boldsymbol w}\\
{\rm subject~ to} ~ {\boldsymbol w}^H{\boldsymbol a}(\theta_k)~ &
=1 \label{flcmv}
\end{split}
\end{equation}
The solution to the problem in (\ref{flcmv}) is given by
\cite{haykin,frost72}
\begin{equation}
{\boldsymbol w}_{\rm opt} =  \frac{ {\boldsymbol R}^{-1}{\boldsymbol
a}(\theta_k)}{ {\boldsymbol a}^H(\theta_k) {\boldsymbol R}^{-1}
{\boldsymbol a}( \theta_k) \big)}, \label{frbf}
\end{equation}
where ${\boldsymbol a}(\theta_k)$ is the steering vector of the SoI,
${\boldsymbol r}(i)$ is the received data, the covariance matrix of
${\boldsymbol r}(i)$ is described by ${\boldsymbol R}=E[{\boldsymbol
r}(i){\boldsymbol r}^H(i)]$, $(\cdot)^{H}$ denotes Hermitian
transpose and $E[\cdot]$ stands for expected value. The beamformer
${\boldsymbol w}(i)$ can be estimated via SG or RLS algorithms
\cite{haykin}. However, the laws that govern their convergence and
tracking behaviors imply that they depend on $M$ and on the
eigenvalue spread of ${\boldsymbol{R}}$.

A reduced-rank algorithm must extract the most important features of
the processed data by performing dimensionality reduction. This
mapping is carried out by a $M \times D$ rank-reduction matrix
${\boldsymbol S}_{D}$ on the received data as given by
\begin{equation}
\bar{\boldsymbol r}(i) = {\boldsymbol S}_D^H {\boldsymbol r}(i)
\end{equation}
where, in what follows, all $D$-dimensional quantities are denoted
with a "bar". The resulting projected received vector
$\bar{\boldsymbol r}(i)$ is the input to a beamformer represented by
the $D \time 1$ vector $\bar{\boldsymbol w}=[ \bar{w}_1^{}
~\bar{w}_2^{}~\ldots\bar{w}_D^{}]^T$. The filter output is
\begin{equation}
\bar{x}(i) = \bar{\boldsymbol w}^{H}\bar{\boldsymbol r}(i)
\end{equation}
In order to design a reduced-rank beamformer $\bar{\boldsymbol w}$
we consider the following optimization problem
\begin{equation}
\begin{split}
{\textrm{minimize}}  ~ E\big[ |\bar{\boldsymbol
w}^{H}\bar{\boldsymbol r}(i)|^2 \big] & = \bar{\boldsymbol
w}^{H} \bar{\boldsymbol R} \bar{\boldsymbol w} \\
{\textrm {subject to}}  ~ \bar{\boldsymbol w}^H\bar{\boldsymbol
a}(\theta_k) & = 1
\end{split}
\end{equation}
The solution to the above problem is
\begin{equation}
\begin{split}
\bar{\boldsymbol w}_{\rm opt} & =  \frac{ \bar{\boldsymbol
R}^{-1}\bar{\boldsymbol a}(\theta_k)}{ \bar{\boldsymbol
a}^H(\theta_k) \bar{\boldsymbol R}^{-1} \bar{\boldsymbol a}(
\theta_k)}\\  & =  \frac{ ({\boldsymbol S}_D^H{\boldsymbol
R}{\boldsymbol S}_D )^{-1}{\boldsymbol S}_D^H{\boldsymbol
a}(\theta_k)}{ {\boldsymbol a}^H{\boldsymbol S}_D(\theta_k)
({\boldsymbol S}_D^H{\boldsymbol R}{\boldsymbol S}_D )^{-1}
{\boldsymbol S}_D^H{\boldsymbol a}( \theta_k)},\label{rrbf}
\end{split}
\end{equation}
where the reduced-rank covariance matrix is $\bar{\boldsymbol R} =
E[ \bar{\boldsymbol r}(i)\bar{\boldsymbol r}^{H}(i)]={\boldsymbol
S}_D^H{\boldsymbol R}{\boldsymbol S}_D$ and the reduced-rank
steering vector is $\bar{\boldsymbol a}(\theta_k)={\boldsymbol
S}_D^H {\boldsymbol a}(\theta_k)$. The above development shows that
the choice of ${\boldsymbol S}_D$ to perform dimensionality
reduction on ${\boldsymbol r}(i)$ is very important, and can lead to
an improved convergence and tracking performance over the full-rank
beamformer. A key problem with the full-rank and the reduced-rank
beamformers described in (\ref{frbf}) and (\ref{rrbf}),
respectively, is that their performance is deteriorated when they
employ the presumed array steering vector ${\boldsymbol
a}_p(\theta_k)$. In these situations it is fundamental to employ a
robust technique that can mitigate the effects of the mismatches
between the actual and the presumed steering vector.

\subsection{Robust Adaptive LCMV Beamformers}

An effective technique for \index{robust beamforming} is the use of
diagonal loading strategies \cite{cox,vorobyov03,lorenz,chen07}. In
what follows, robust full-rank and reduced-rank LCMV beamforming
designs are described. A general approach based on diagonal loading
is employed for both full-rank and reduced-rank designs.

A robust full-rank LCMV beamformer represented by an $M \times 1$
vector ${\boldsymbol w}$ can be designed by solving the following
optimization problem
\begin{equation}
\begin{split}
{\rm minimize} ~  E[|{\boldsymbol w}^H {\bf r}(i) |^2] + \epsilon^2 ||{\boldsymbol w}||^2 & = {\boldsymbol w}^{H} {\boldsymbol R} {\boldsymbol w}+ \epsilon^2 {\boldsymbol w}^H{\boldsymbol w}\\
{\rm subject~ to} ~ {\boldsymbol w}^H{\boldsymbol a}(\theta_k)~ & =
1 , \label{rflcmv}
\end{split}
\end{equation}
where $\epsilon^2$ is constant that needs to be chosen by the
designer. The solution to the problem in (\ref{rflcmv}) is given by
\begin{equation}
{\boldsymbol w}_{\rm opt} =  \frac{ ({\boldsymbol R}+\epsilon^2
{\boldsymbol I}_M)^{-1}{\boldsymbol a}_p(\theta_k)}{ {\boldsymbol
a}_p^H(\theta_k) ({\boldsymbol R}+\epsilon^2 {\boldsymbol I}_M)^{-1}
{\boldsymbol a}_p( \theta_k) \big)}
\end{equation}
where ${\boldsymbol a}_p(\theta_k)$ is the presumed steering vector
of the SoI and ${\boldsymbol I}_D$ is an $M$-dimensional identity
matrix. It turns out that the adjustment of $\epsilon^2$ needs to be
obtained numerically by an optimization algorithm.

In order to design a robust reduced-rank LCMV beamformer
$\bar{\boldsymbol w}$ we follow a similar approach to the full-rank
case and consider the following optimization problem
\begin{equation}
\begin{split}
{\textrm{minimize}}  ~ E\big[ |\bar{\boldsymbol w}^{H}{\boldsymbol
S}_D^H{\boldsymbol r}(i)|^2 \big] + \epsilon^2 ||{\boldsymbol
S}_D\bar{\boldsymbol w}||^2 & = \bar{\boldsymbol w}^{H} {\boldsymbol
S}_D^H {\boldsymbol R}{\boldsymbol S}_D \bar{\boldsymbol w} \\ &
\quad + \epsilon^2 \bar{\boldsymbol w}^H{\boldsymbol S}_D^H
{\boldsymbol
S}_D\bar{\boldsymbol w} \\
{\textrm {subject to}}  ~ \bar{\boldsymbol w}^H{\boldsymbol
S}_D^H{\boldsymbol a}_p(\theta_k) & = 1 , \label{rrrbf}
\end{split}
\end{equation}
The solution to the above problem is
\begin{equation}
\bar{\boldsymbol w}_{\rm opt} =  \frac{ ({\boldsymbol
S}_D^H{\boldsymbol R}{\boldsymbol S}_D +\epsilon^2  {\boldsymbol
I}_D)^{-1}{\boldsymbol S}_D^H{\boldsymbol a}_p(\theta_k)}{
{\boldsymbol a}^H_p{\boldsymbol S}_D(\theta_k) ({\boldsymbol
S}_D^H{\boldsymbol R}{\boldsymbol S}_D +\epsilon^2  {\boldsymbol
I}_D)^{-1} {\boldsymbol S}_D^H{\boldsymbol a}_p( \theta_k)}
\end{equation}
where the tuning of $\epsilon^2$ requires an algorithmic approach as
there is no closed-form solution and ${\boldsymbol I}_D$ is a
$D$-dimensional identity matrix.

\section{Robust Reduced-Rank Beamforming Based on Joint Iterative Optimization of Parameters}

In this section, the principles of the robust reduced-rank
beamforming scheme based on joint iterative optimization of
parameters, termed RJIO, is introduced. The RJIO scheme, depicted in
Fig. 2, employs a rank-reduction matrix ${\boldsymbol S}_{D}(i)$
with dimensions $M \times D$ to perform dimensionality reduction on
a data vector ${\boldsymbol r}(i)$ with dimensions $M \times 1$. The
reduced-rank beamformer $\bar{\boldsymbol w}(i)$ with dimensions $D
\times 1$ processes the reduced-rank data vector $\bar{\boldsymbol
r}(i)$ in order to yield a scalar estimate $\bar{x}(i)$. The
rank-reduction matrix ${\bf S}_{D}(i)$ and the reduced-rank
beamformer $\bar{\boldsymbol w}(i)$ are jointly optimized in the
RJIO scheme according to the MV criterion subject to a robust
constraint that ensures that the beamforming algorithm is robust
against steering vector mismatches and short data records.

\begin{figure}[h]
       \vspace*{-0.0em}\centering  % figura centralizada
       \hspace*{-0.5em}{\includegraphics[width=10.0cm, height=4.2cm]{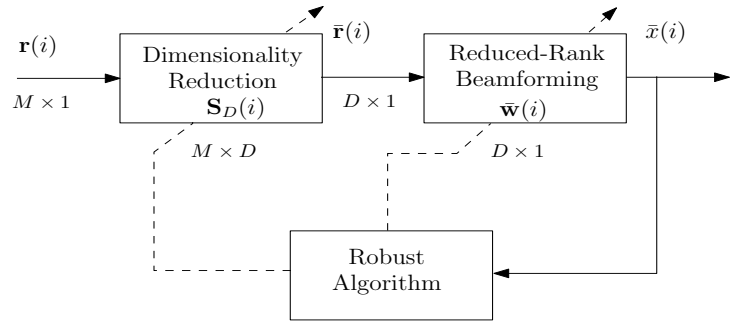}}
        \vspace*{-1em}
       \caption{Block diagram of the RJIO scheme}
 \label{SchemeSetup}
\end{figure}

In order to describe the RJIO method, let us first consider the
structure of the $M \times D$ rank-reduction matrix
\begin{equation}
{\boldsymbol S}_{D}(i) = [~{\boldsymbol s}_1(i) ~| ~{\boldsymbol
s}_2(i)~| ~\ldots~|{\boldsymbol s}_D(i)~]
\end{equation}
where the columns ${\boldsymbol s}_d(i)$ for $d = 1,~\ldots,~D$
constitute a bank of $D$ robust beamformers with dimensions $M
\times 1$ as given by
\begin{equation}
{\boldsymbol s}_d(i)=[s_{1,d}(i) ~ s_{2,d}(i)~ \ldots~s_{M,d}(i)
]^T \nonumber
\end{equation}
The output $\bar{x}(i)$ of the RJIO scheme can be expressed as a
function of the input vector ${\bf r}(i)$, the matrix ${\bf S}_D(i)$
and the reduced-rank beamformer $\bar{\bf w}(i)$:
\begin{equation}
\begin{split}
\bar{x}(i) & =  \bar{\boldsymbol w}^H(i) {\boldsymbol S}_D^H(i)
{\boldsymbol r}(i) = \bar{\boldsymbol w}^H(i) \bar{\boldsymbol
r}(i)
\end{split}
\end{equation}
It is interesting to note that for $D=1$, the RJIO scheme becomes a
robust full-rank LCMV beamforming scheme with an addition weight
parameter $w_D$ that provides an amplitude gain. For $D>1$, the
signal processing tasks are changed and the robust full-rank LCMV
beamformers compute a subspace projection and the reduced-rank
beamformer provides a unity gain in the direction of the SoI. This
rationale is fundamental to the exploitation of the low-rank nature
of signals in typical beamforming scenarios.

The robust LCMV expressions for ${\boldsymbol S}_D(i)$ and
$\bar{\boldsymbol w}(i)$ can be computed via the following
optimization problem
\begin{equation}
\begin{split}
{\textrm{minimize}} ~ & E\big[ |\bar{\boldsymbol
w}^{H}(i){\boldsymbol S}_D^H(i){\boldsymbol r}(i)|^2 \big] +
\epsilon^2 ||{\boldsymbol S}_D(i)\bar{\boldsymbol w}(i)||^2 =\\
 & \bar{\boldsymbol w}^{H}(i) {\boldsymbol S}_D^H(i) {\boldsymbol
R} {\boldsymbol S}_D(i) \bar{\boldsymbol w}(i)  +
\epsilon^2 \bar{\boldsymbol w}^H(i){\boldsymbol S}_D^H(i) {\boldsymbol S}_D(i)\bar{\boldsymbol w}(i)\\
{\textrm {subject to}}  ~&  \bar{\boldsymbol w}^H(i){\boldsymbol
S}_D^H(i) {\boldsymbol a}_p(\theta_k)  = 1 \label{propt}
\end{split}
\end{equation}
In order to solve the above problem, we resort to the method of
Lagrange multipliers \cite{haykin} and transform the
\index{constrained optimization} into an unconstrained one expressed
by the Lagrangian
\begin{equation}
\begin{split}
{\mathcal L}({\boldsymbol S}_D(i), \bar{\boldsymbol
w}(i),\epsilon^2(i)) & = E\big[ |\bar{\boldsymbol
w}^{H}(i){\boldsymbol S}_D^H(i){\boldsymbol r}(i)|^2 \big] \\ &
\quad + \epsilon^2(i) \bar{\boldsymbol w}^H(i){\boldsymbol
S}_D^H(i){\boldsymbol S}_D(i)\bar{\boldsymbol w}(i)) \\ & \quad +
[\lambda ( \bar{\boldsymbol w}^H(i){\bf S}_D^H(i){\boldsymbol
a}_p(\theta_k)- 1 ], \label{uopt}
\end{split}
\end{equation}
where $\lambda$ is a scalar Lagrange multiplier, $*$ denotes complex
conjugate. By fixing $\bar{\boldsymbol w}(i)$, minimizing
(\ref{uopt}) with respect to ${\boldsymbol S}_D(i)$ and solving for
$\lambda$, we get
\begin{equation}
\begin{split}
{\boldsymbol S}_D(i) & = \frac{ ({\boldsymbol R}+\epsilon^2(i)
{\boldsymbol I}_M)^{-1} {\boldsymbol a}_p(\theta_k) \bar{\boldsymbol
w}^H(i) \bar{\boldsymbol R}_{\bar{w}}^{-1}}{\bar{\boldsymbol
w}^H(i)\bar{\boldsymbol R}_{\bar{w}}^{-1}\bar{\boldsymbol w}(i)
{\boldsymbol a}^H_p(\theta_k) ({\boldsymbol R}(i)+\epsilon^2(i)
{\boldsymbol I}_M)^{-1} {\boldsymbol a}_p(\theta_k)}, \label{filts}
\end{split}
\end{equation}
where  ${\boldsymbol R} = E[{\boldsymbol r}(i){\boldsymbol
r}^{H}(i)]$ and $\bar{\boldsymbol R}_{\bar{w}} = E[\bar{\boldsymbol
w}(i)\bar{\boldsymbol w}^{H}(i)]$. By fixing ${\boldsymbol S}_D(i)$,
minimizing (\ref{uopt}) with respect to $\bar{\boldsymbol w}(i)$ and
solving for $\lambda$, we arrive at the expression
\begin{equation}
\bar{\boldsymbol w}(i) =  \frac{ (\bar{\boldsymbol R}(i)+
\epsilon^2(i) {\boldsymbol S}_D^H(i){\boldsymbol I}_D{\boldsymbol
S}_D(i))^{-1} \bar{\boldsymbol a}_p(\theta_k)}{\bar{\boldsymbol
a}^H_p(\theta_k)(\bar{\boldsymbol R}(i)+ \epsilon^2(i) {\boldsymbol
S}_D^H(i){\boldsymbol I}_D{\boldsymbol S}_D(i))^{-1}\bar{\boldsymbol
a}_p(\theta_k)}, \label{filtw}
\end{equation}
where $\bar{\boldsymbol R}(i) = E[{\boldsymbol S}_D^H(i){\boldsymbol
r}(i){\boldsymbol r}^H(i)  {\boldsymbol S}_D(i)] =E[\bar{\boldsymbol
r}(i) \bar{\boldsymbol r}^{H}(i)]$, $\bar{\boldsymbol a}_p(\theta_k)
= {\boldsymbol S}_D^H(i){\boldsymbol a}_p(\theta_k)$. Note that the
filter expressions in (\ref{filts}) and (\ref{filtw}) are not
closed-form solutions for $\bar{\boldsymbol w}(i)$ and ${\boldsymbol
S}_D(i)$ since (\ref{filts}) is a function of $\bar{\boldsymbol
w}(i)$ and (\ref{filtw}) depends on ${\boldsymbol S}_D(i)$. Thus, it
is necessary to iterate (\ref{filts}) and (\ref{filtw}) with initial
values to obtain a solution. The key strategy lies in the robust
joint optimization of the beamformers. The rank $D$ and the diagonal
loading parameter $\epsilon^2(i)$ must be adjusted by the designer
to ensure appropriate performance or can be estimated via another
algorithm. In the next section, iterative solutions via adaptive
algorithms are sought for the robust computation of ${\boldsymbol
S}_D(i)$, $\bar{\boldsymbol w}(i)$, the diagonal loading
$\epsilon(i)$ and the rank adaptation.

\section{Adaptive Algorithms}
\label{sec:typestyle}

In this section, adaptive SG and RLS versions of the RJIO scheme are
developed for an efficient implementation. The important issue of
determining the rank of the scheme with an adaptation technique is
considered. The computational complexity in arithmetic operations of
the RJIO-based algorithms is then detailed.

\subsection{Stochastic Gradient Algorithm}

In this part, we present a low-complexity SG adaptive reduced-rank
algorithm for an efficient implementation of the RJIO method. The
basic idea is to employ an alternating optimization strategy to
update ${\boldsymbol S}_D(i)$, $\bar{\boldsymbol w}(i)$ and the
diagonal loading $\epsilon^2(i)$ By computing the instantaneous
gradient terms of (\ref{uopt}) with respect to ${\boldsymbol
S}_D(i)$, $\bar{\boldsymbol w}(i)$ and $\epsilon^2(i)$, we obtain
\begin{equation}
\begin{split}
\nabla {{\mathcal L}_{MV}}_{{\boldsymbol S}_D^*(i)} & = \bar{x}^*(i)
{\boldsymbol r}(i)\bar{\boldsymbol w}^H(i) +
\epsilon^2(i){\boldsymbol
S}_D(i)\bar{\boldsymbol w}(i)\bar{\boldsymbol w}^H(i) + 2 \lambda^* {\bf a}_p(\theta_k) \bar{\boldsymbol w}^H(i) \\
\nabla {{\mathcal L}_{MV}}_{\bar{\boldsymbol w}^*(i)} & =
\bar{x}^*(i) {\boldsymbol S}_D^H(i){\bf r}(i) +
\epsilon^2(i){\boldsymbol S}_D^H(i){\boldsymbol
S}_D(i)\bar{\boldsymbol w}(i) + 2 \lambda^* {\boldsymbol
S}_D^H(i){\bf a}_p(\theta_k)\\
\nabla {{\mathcal L}_{MV}}_{\epsilon^2(i)} & = 2\epsilon(i)
{\boldsymbol w}^H(i){\boldsymbol S}_D^H(i){\boldsymbol
S}_D(i)\bar{\boldsymbol w}(i)
\end{split}
\end{equation}
By introducing the positive step sizes $\mu_s$, $\mu_w$ and
$\mu_{\epsilon}$, using the gradient rules ${\boldsymbol S}_D(i+1) =
{\boldsymbol S}_D(i) - \mu_s \nabla {{\mathcal
L}_{MV}}_{{\boldsymbol S}_D^*(i)}$, $\bar{\boldsymbol w}(i+1) =
\bar{\boldsymbol w}(i) - \mu_w \nabla {{\mathcal
L}_{MV}}_{\bar{\boldsymbol w}^*(i)}$ and $\epsilon(i+1) =
\epsilon(i) - \mu_w \nabla {{\mathcal L}_{MV}}_{\epsilon(i)}$,
enforcing the constraint and solving the resulting equations, we
obtain
\begin{equation}
\begin{split}
{\boldsymbol S}_D(i+1) & = {\boldsymbol S}_D(i) - \mu_s \big[
\bar{x}^*(i){\boldsymbol r}(i)\bar{\boldsymbol w}^H(i)+ \epsilon(i)
{\boldsymbol S}_D(i)\bar{\boldsymbol w}(i)\bar{\boldsymbol w}^H(i)\\
& \quad - \big({\boldsymbol a}^H_p(\theta_k){\boldsymbol
a}_p(\theta_k)\big)^{-1}{\boldsymbol a}_p(\theta_k) \bar{\boldsymbol
w}^H(i) (\bar{x}^*(i){\boldsymbol a}^H_p (\theta_k) {\boldsymbol
r}(i)+ \epsilon(i))\big], \label{recsd}
\end{split}
\end{equation}
\begin{equation}
\begin{split}
\bar{\boldsymbol w}(i+1) & = \bar{\boldsymbol w}(i) - \mu_w \big(
\bar{x}^*(i) {\boldsymbol S}_D^H(i){\boldsymbol r}(i) +
\epsilon(i){\boldsymbol S}_D^H(i){\boldsymbol S}_D(i)
\bar{\boldsymbol w}(i) \\ & \quad + ({\boldsymbol
a}_p^H(\theta_k){\boldsymbol a}_p(\theta_k))^{-1} ( \bar{x}^*(i)
{\boldsymbol r}^H(i) {\boldsymbol S}_D(i) {\boldsymbol S}_D^H(i)
{\boldsymbol a}_p(\theta_k)  + \epsilon(i){\boldsymbol w}^H(i)
{\boldsymbol S}_D^H(i) {\boldsymbol S}_D(i){\boldsymbol
S}_D^H(i){\boldsymbol a}_p(\theta_k) \big) \label{recw},
\end{split}
\end{equation}
\begin{equation}
\begin{split}
\epsilon(i+1) & = \epsilon(i) - \mu_{\epsilon} \bar{\boldsymbol
w}^H(i) {\boldsymbol S}_D^H(i)  {\boldsymbol S}_D(i)\bar{\boldsymbol
w}(i) \label{rece},
\end{split}
\end{equation}
where $\bar{x}(i)= \bar{\boldsymbol w}^H(i) {\boldsymbol S}_D^H(i)
{\boldsymbol r}(i)$. The RJIO scheme trades-off a full-rank
beamformer against one rank-reduction matrix ${\boldsymbol S}_D(i)$,
one reduced-rank beamformer $\bar{\boldsymbol w}(i)$ and one
adaptive loading recursion operating in an alternating fashion and
exchanging information.

\subsection{Recursive Least Squares Algorithms}

Here, an RLS algorithm is devised for an efficient implementation of
the RJIO method. To this end, let us first consider the Lagrangian
\begin{equation}
\begin{split}
\label{costls} {\mathcal{L}}_{\rm LS}({\boldsymbol S}_D(i),
\bar{\boldsymbol w}(i), \epsilon(i)) &
=\sum_{l=1}^{i}\alpha^{i-l}\big|\bar{\boldsymbol
w}^{H}(i){\boldsymbol S}_{D}^{H}(i)\boldsymbol r(l)\big|^{2} +
\epsilon^2(i) \bar{\boldsymbol w}^H(i) {\boldsymbol S}_D^H(i)
 {\boldsymbol
S}_D(i)\bar{\boldsymbol w}(i) + \lambda\big(\bar{\boldsymbol
w}^{H}(i) {\boldsymbol S}_{D}^{H}(i){\boldsymbol
a}_p(\theta_k)-1\big)
\end{split}
\end{equation}
where $\alpha$ is the forgetting factor chosen as a positive
constant close to, but less than $1$.

Fixing $\bar{\boldsymbol w}(i)$, computing the gradient of
(\ref{costls}) with respect to $\boldsymbol S_{D}(i)$, equating the
gradient terms to zero and solving for $\lambda$, we obtain
\begin{equation}\label{18}
{\boldsymbol S}_{D}(i)=\frac{\boldsymbol P(i)\boldsymbol
a_p(\theta_k){\boldsymbol a}^{H}_p(\theta_k) {\boldsymbol
S}_D(i-1)}{\boldsymbol a^{H}_p(\theta_k)\boldsymbol P(i)\boldsymbol
a_p(\theta_k)}
\end{equation}
where we defined the inverse covariance matrix ${\boldsymbol P}(i)
= \boldsymbol R^{-1}(i)$ for convenience of presentation.
Employing the matrix inversion lemma \cite{haykin}, we obtain
\begin{equation}\label{19}
\boldsymbol k(i)=\frac{\alpha^{-1}\boldsymbol P(i-1)\boldsymbol
r(i)}{1+\alpha^{-1}\boldsymbol r^{H}(i)\boldsymbol
P(i-1)\boldsymbol r(i)}
\end{equation}
\begin{equation}\label{20}
\boldsymbol P(i)=\alpha^{-1}\boldsymbol
P(i-1)-\alpha^{-1}\boldsymbol k(i)\boldsymbol r^{H}(i)\boldsymbol
P(i-1) + \epsilon^2(i){\boldsymbol I}_M
\end{equation}
where $\boldsymbol k(i)$ is the $M \times 1$ Kalman gain vector. We
set $\boldsymbol P(0)=\delta\boldsymbol I_M$ to start the recursion
of (\ref{20}), where $\delta$ is a positive constant.

Assuming $\boldsymbol S_{D}(i)$ is known and taking the gradient of
(\ref{costls}) with respect to $\bar{\boldsymbol w}(i)$, equating
the terms to a null vector and solving for $\lambda$, we obtain the
$D \times 1$ reduced-rank beamformer
\begin{equation}\label{13}
\bar{\boldsymbol w}(i)=\frac{\bar{\boldsymbol P}(i){\boldsymbol
S}_{D}^H(i){\boldsymbol a}_p(\theta_k)}{{\boldsymbol
a}^{H}_p(\theta_k){\boldsymbol S}_{D}(i)\bar{\boldsymbol
P}(i){\boldsymbol S}_{D}^H(i){\boldsymbol a}_p(\theta_k)}
\end{equation}
where $\bar{\boldsymbol P}(i) = \bar{\boldsymbol R}^{-1}(i)$ and
$\bar{\boldsymbol R}(i)=\sum_{l=1}^{i}\alpha^{i-l}\bar{\boldsymbol
r}(l)\bar{\boldsymbol r}^{H}(l)$ is the reduced-rank input
covariance matrix. In order to estimate $\bar{\boldsymbol P}(i)$,
we use the matrix inversion lemma \cite{haykin} as follows
\begin{equation}\label{15}
\bar{\boldsymbol k}(i)=\frac{\alpha^{-1}\bar{\boldsymbol
P}(i-1)\bar{\boldsymbol r}(i)}{1+\alpha^{-1}\bar{\boldsymbol
r}^{H}(i)\bar{\boldsymbol P}(i-1)\bar{\boldsymbol r}(i)}
\end{equation}
\begin{equation}\label{16}
\bar{\boldsymbol P}(i)=\alpha^{-1}\bar{\boldsymbol
P}(i-1)-\alpha^{-1}\bar{\boldsymbol k}(i)\bar{\boldsymbol
r}^{H}(i)\bar{\boldsymbol P}(i-1) + \epsilon^2(i){\boldsymbol I}_D
\end{equation}
where $\bar{\boldsymbol k}(i)$ is the $D \times 1$ reduced-rank gain
vector and $\bar{\boldsymbol P}(i)=\bar{\boldsymbol R}^{-1}(i)$ is
referred to as the reduced-rank inverse covariance matrix. Hence,
the covariance matrix inversion $\bar{\boldsymbol R}^{-1}(i)$ is
replaced at each step by the recursive processes (\ref{15}) and
(\ref{16}) for reducing the complexity.  The recursion of (\ref{16})
is initialized by choosing $\bar{\boldsymbol
P}(0)=\bar{\delta}{\boldsymbol I}_D$, where $\bar{\delta}$ is a
positive constant. The last recursion adjusts the diagonal loading
according to the following update equation
\begin{equation}
\begin{split}
\epsilon(i+1) & = \epsilon(i) - \mu_{\epsilon} \bar{\boldsymbol
w}^H(i) {\boldsymbol S}_D^H(i)  {\boldsymbol S}_D(i)\bar{\boldsymbol
w}(i) \label{rece},
\end{split}
\end{equation}
The RJIO-RLS algorithm trades-off a full-rank beamformer with $M$
coefficients against one matrix recursion to compute ${\boldsymbol
S}_D(i)$, given in (\ref{18})-(\ref{20}), one $ D \times 1$
reduced-rank adaptive beamformer $\bar{\boldsymbol w}(i)$, given in
(\ref{13})-(\ref{16}), and one recursion to adjust the diagonal
loading described in (\ref{rece}) in an alternating manner and
exchanging information.

\subsection{Complexity of RJIO Algorithms}

Here, we evaluate the computational complexity of the RJIO and
analyzed LCMV algorithms. The complexity expressed in terms of
additions and multiplications is depicted in Table I. We can verify
that the RJIO-SG algorithm has a complexity that grows linearly with
$DM$, which is about $D$ times higher than the full-rank LCMV-SG
algorithm and significantly lower than the remaining techniques. If
$D << M$ (as we will see later) then the additional complexity can
be acceptable provided the gains in performance justify them. In the
case of the RJIO-RLS algorithm the complexity is quadratic with
$M^2$ and $D^2$. This corresponds to a complexity slightly higher
than the one observed for the full-rank LCMV-RLS algorithm, provided
$D$ is significantly less than $M$, and lower than the \index{robust
beamforming} algorithms WC-SOC \cite{vorobyov03} and WC-ME
\cite{li03}.

\begin{table}[h]
\centering%
\caption{\small Computational complexity of LCMV algorithms} {
\begin{tabular}{lcc}
\hline
{\small Algorithm} & {\small Additions} & {\small Multiplications} \\
\hline \emph{\small \bf LCMV-SG }\cite{frost72}  & {\small
$3M+1$} & {\small $3M+2$} \\ \\
\emph{\small \bf LCMV-RLS }\cite{romano96} & {\small $3M^2 - 2M +
3$} & {\small $6M^{2}+2M + 2$}
\\ \\
\emph{\small \bf RJIO-SG}   & {\small $3DM + 4M $}  & {\small $ 5DM + 2M $}  \\
\emph{\small \bf }  & {\small $ +2D-2$}  & {\small $  +5D + 2$}
\\\\
\emph{\small \bf RJIO-RLS}  & {\small $ 3M^2 - M + 3 $}  & {\small $ 7M^2 + 3M $}  \\
\emph{\small \bf }  & {\small $+3D^2 - 7D + 3$}  & {\small $ +7D^2+10D$}
\\ \\
\emph{\small \bf SMI }\cite{goldstein} & {\small $2/3M^3 + 3M^2 $} &
{\small $2/3M^3 + 5M^2 $}  \\
%\emph{\small \bf MSWF-RLS }\cite{goldstein} & {\small $DM^2 + M^2
%+ 6D^2 $} & {\small $DM^2+M^2$} \\
%\emph{\small } & {\small $-8D+2$} & {\small $+2DM+3D+2$} \\
%\emph{\small \bf AVF }\cite{qian} & {\small $D((M)^2+3(M-1)^2)-1$} & {\small $D(4M^2+4M + 1)$} \\
%\emph{\small \bf  } & {\small $+D(5(M-1)+1)+2M$} & {\small $+4M +
%2$}
\\\hline
\end{tabular}
}
\end{table}

In order to illustrate the main trends in what concerns the
complexity of the proposed and analyzed algorithms, we show in Fig.
3 the complexity in terms of additions and multiplications versus
the number of input samples $M$. The curves indicate that the
RJIO-RLS algorithm has a complexity lower than the WC-ME \cite{li03}
and the WC-SOC \cite{vorobyov03}, whereas it remains at the same
level of the full-rank LCMV-RLS algorithm. The RJIO-SG algorithm has
a complexity that is situated between the full-rank LCMV-RLS and the
full-rank LCMV-SG algorithms.

\begin{figure}[!htb]
\begin{center}
\def\epsfsize#1#2{1\columnwidth}
\epsfbox{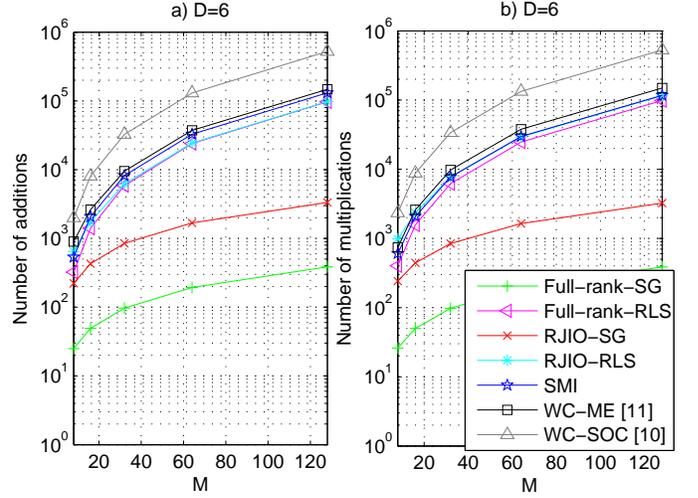} \vspace*{-1.0em} \caption{Computational
complexity in terms of arithmetic operations against $M$ }
\end{center}
\end{figure}

\subsection{Rank Adaptation }

The performance of the algorithms described in the previous
subsections depends on the rank $D$. This motivates the development
of methods to automatically adjust $D$ on the basis of the cost
function. Differently from existing methods for rank adaptation
which use MSWF-based algorithms \cite{goldstein} or AVF-based
recursions \cite{qian}, we focus on an approach that jointly
determines $D$ based on the LS criterion computed by the filters
${\boldsymbol S}_D(i)$ and $\bar{\boldsymbol w}_D(i)$, where the
subscript $D$ denotes the rank used for the adaptation. In
particular, we present a method for automatically selecting the
ranks of the algorithms based on the exponentially weighted
\textit{a posteriori} least-squares type cost function described by
\begin{equation}
{\mathcal C}({\boldsymbol S}_D(i-1),\bar{\boldsymbol w}_D(i-1)) =
\sum_{l=1}^{i} \alpha^{i-l} \big|\bar{\boldsymbol
w}_D^{H}(i-1){\boldsymbol S}_D(i-1){\boldsymbol r}(l)|^2 ,
\label{eq:costadap}
\end{equation}
where $\alpha$ is the forgetting factor and $\bar{\bf w}_D(i-1)$ is
the \index{reduced-rank} beamformer with rank $D$. For each time
interval $i$, we can select the rank $D_{\rm opt}$ which minimizes
${\mathcal C}({\boldsymbol S}_D(i-1),\bar{\boldsymbol w}_{D}(i-1))$
and the exponential weighting factor $\alpha$ is required as the
optimal rank varies as a function of the data record. The key
quantities to be updated are the rank-reduction matrix ${\boldsymbol
S}_D(i)$, the \index{reduced-rank} beamformer $\bar{\boldsymbol
w}_D(i)$, the associated presumed reduced-rank steering vector
$\bar{\boldsymbol a}_p(\theta_k)$ and the inverse of the
reduced-rank covariance matrix $\bar{\boldsymbol P}(i)$ (for the
RJIO-RLS algorithm). To this end, we define the following extended
rank-reduction matrix $ {\boldsymbol S}_{D}(i)$ and the extended
reduced-rank beamformer weight vector $\bar{\boldsymbol w}_{D}(i)$
as follows:
\begin{equation}
{\boldsymbol S}_{D}(i) = \left[\begin{array}{cccccc} s_{1,1} &
s_{1,2} & \ldots & s_{1,D_{\rm min}} & \ldots & s_{1,D_{\rm max}}
\\ \vdots & \vdots & \vdots & \vdots & \ddots & \vdots \\
s_{M,1} & s_{M,2} & \ldots & s_{M,D_{\rm min}} & \ldots &
s_{M,D_{\rm max}} \end{array} \right] ~~{\rm and} ~~
\bar{\boldsymbol w}_{D}(i) = \left[\begin{array}{c} w_1 \\ w_2
\\ \vdots \\ w_{D_{\rm min}} \\ \vdots \\ w_{D_{\rm max}} \end{array}\right]
\end{equation}
The extended rank-reduction matrix $ {\boldsymbol S}_{D}(i)$ and the
extended reduced-rank beamformer weight vector $\bar{\boldsymbol
w}_{D}(i)$ are updated along with the associated quantities
$\bar{\boldsymbol a}(\theta_k)$ and $\bar{\boldsymbol P}(i)$ (only
for the RLS) for the maximum allowed rank $D_{\rm max}$ and then the
rank adaptation algorithm determines the rank that is best for each
time instant $i$ using the cost function in (\ref{eq:costadap}). The
rank adaptation algorithm is then given by
\begin{equation}
D_{\rm opt} = \arg \min_{D_{\rm min} \leq d \leq D_{\rm max}}
{\mathcal C}({\boldsymbol S}_D(i-1),\bar{\boldsymbol w}_D(i-1))
\end{equation}
where $d$ is an integer, $D_{\rm min}$ and $D_{\rm max}$ are the
minimum and maximum ranks allowed for the reduced-rank beamformer,
respectively. Note that a smaller rank may provide faster adaptation
during the initial stages of the estimation procedure and a greater
rank usually yields a better steady-state performance. Our studies
reveal that the range for which the rank $D$ of the proposed
algorithms have a positive impact on the performance of the
algorithms is limited, being from $D_{\rm min}=3$ to $D_{\rm max}=8$
for the reduced-rank beamformer recursions. These values are rather
insensitive to the system load (number of users), to the number of
array elements and work very well for all scenarios and algorithms
examined. The additional complexity of the proposed rank adaptation
algorithm is that it requires the update of all involved quantities
with the maximum allowed rank $D_{\rm max}$ and the computation of
the cost function in (\ref{eq:costadap}). This procedure can
significantly improve the convergence performance and can be relaxed
(the rank can be made fixed) once the algorithm reaches steady
state. Choosing an inadequate rank for adaptation may lead to
performance degradation, which gradually increases as the adaptation
rank deviates from the optimal rank.

\section{Simulations}

In this section, the performance of the RJIO and some existing
beamforming algorithms is assessed using computer simulations. A
sensor-array system with a ULA equipped with $M$ sensor elements is
considered for assessing the beamforming algorithms. In particular,
the performance of the RJIO scheme with SG and RLS algorithms is
compared with existing techniques, namely, the full-rank LCMV-SG
\cite{frost72} and LCMV-RLS \cite{romano96}, and the robust
techniques reported in \cite{vorobyov03}, termed WC-SOC, and
\cite{li03}, called Robust-ME, and the optimal linear beamformer
that assumes the knowledge of the covariance matrix and the actual
steering vector \cite{li}. In particular, the algorithms are
compared in terms of the signal-to-interference-plus-noise ratio
(SINR), which is defined for the \index{reduced-rank} schemes as
\begin {equation}
\centering {\rm SINR}(i)=\frac{\bar{\boldsymbol
w}^{H}(i){\boldsymbol S}_D^H(i){\boldsymbol R}_{s}{\boldsymbol
S}_D(i) \bar{\boldsymbol w}(i)}{\bar{\boldsymbol
w}^{H}(i){\boldsymbol S}_D^H(i){\boldsymbol R}_{I}{\boldsymbol
S}_D(i)\bar{\boldsymbol w}(i)},
\end{equation}
where ${\boldsymbol R}_{s}$ is the covariance matrix of the desired
signal and ${\boldsymbol R}_{I}$ is the covariance matrix of the
interference and noise in the environment. Note that for the
full-rank schemes the ${\rm SINR}(i)$ assumes ${\boldsymbol
S}_D^H(i) = {\boldsymbol I}_M$. For each scenario, $200$ runs are
used to obtain the curves. In all simulations, the desired signal
power is $\sigma_{d}^{2}=1$, and the signal-to-noise ratio (SNR) is
defined as ${\rm SNR} =\frac{\sigma_d^2}{\sigma^2}$. The beamformers
are initialized as $\bar{\boldsymbol w}(0) = [1 ~ 0~ \ldots ~ 0 ]$
and ${\boldsymbol S}_D(0) = [ {\boldsymbol I}_D^T ~{\boldsymbol
0}_{D \times (M-D)}^T ]$, where ${\boldsymbol 0}_{D \times M-D}$ is
a $D \times (M-D)$ matrix with zeros in all experiments.

In order to assess the performance of the RJIO and other existing
algorithms in the presence of uncertainties, we consider that the
array steering vector is corrupted by local coherent scattering
\begin{align}
{\boldsymbol a}_p(\theta_k)={\boldsymbol a}(\theta_k)+\sum^{4}_{k=1}
\mathrm e^{j \Phi_{k}} {\boldsymbol
a}_{\textrm{sc}}\left(\theta_{k}\right),
\end{align}
where $\Phi_{k}$ is uniformly distributed between zero and $2 \pi$
and $\theta_{k}$ is uniformly distributed with a standard deviation
of $2$ degrees with the assumed direction as the mean. The mismatch
changes for every realization and is fixed over the snapshots of
each simulation trial. In the first two experiments, we consider a
scenario with $7$ interferers at $-60^{o}$, $-45^{o}$, $30^o$
$-15^o$, $0^{o}$, $45^{o}$, $60^{o}$ with powers following a
log-normal distribution with associated standard deviation $3$ dB
around the SoI's power level. The SoI impinges on the array at
$30^o$. The parameters of the algorithms are optimized.

\begin{figure}[!htb]
\begin{center}
\def\epsfsize#1#2{1\columnwidth}
\epsfbox{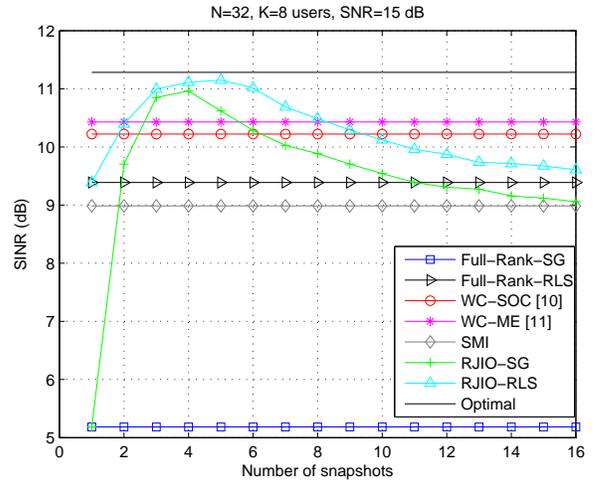} \vspace*{-1em}\caption{\small SINR performance of
LCMV algorithms against rank ($D$) with $M=32$, $SNR=15$ dB, $N=
250$ snapshots} \label{rank}
\end{center}
\end{figure}

We first evaluate the SINR performance of the analyzed algorithms
against the rank $D$ using optimized parameters ($\mu_s$, $\mu_w$
and forgetting factors $\lambda$) for all schemes and $N=250$
snapshots. The results in Fig. \ref{rank} indicate that the best
rank for the RJIO scheme is $D=4$ (which will be used in the second
scenario) and it is very close to the optimal full-rank LCMV
beamformer that has knowledge about the actual steering vector. An
examination of systems with different sizes has shown that $D$ is
relatively invariant to the system size, which brings considerable
computational savings. In practice, the rank $D$ can be adapted in
order to obtain fast convergence and ensure good steady-state
performance and tracking after convergence.

\begin{figure}[!htb]
\begin{center}
\def\epsfsize#1#2{1\columnwidth}
\epsfbox{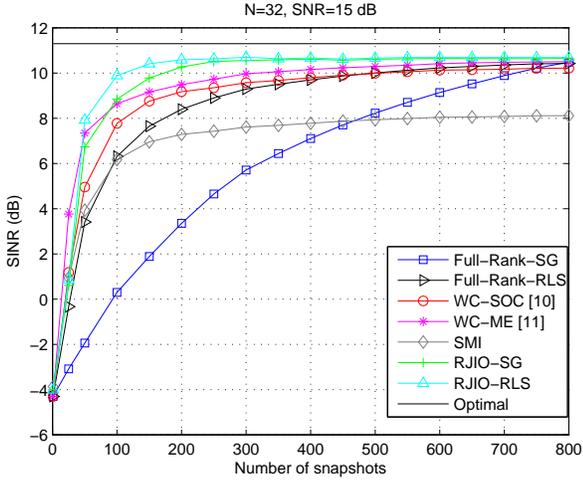} \vspace*{-1em}\caption{\small SINR performance of
robust LCMV algorithms against snapshots with $M=32$, $SNR=15$
dB}\label{snap1}
\end{center}
\end{figure}

We display another scenario in Fig. \ref{snap1} where the robust
adaptive LCMV beamformers are set to converge to the same level of
SINR. The parameters used to obtain these curves are also shown. The
curves show an excellent performance for the RJIO scheme which
converges much faster than the full-rank-SG algorithm, and is also
better than the more complex WC-SOC \cite{vorobyov03} and Robust-ME
\cite{li03} schemes.

\begin{figure}[!htb]
\begin{center}
\def\epsfsize#1#2{1\columnwidth}
\epsfbox{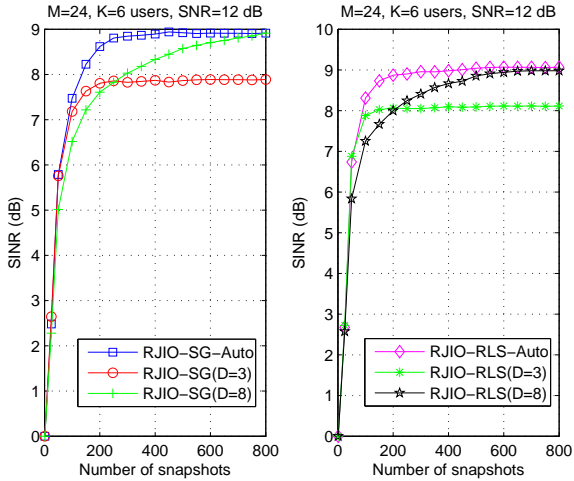} \vspace*{-1em} \caption{SINR performance of
RJIO-LCMV (a) SG and (b) RLS algorithms against snapshots with
$M=24$, $SNR=12$ dB with rank adaptation} \label{auto}
\end{center}
\end{figure}

In the next example, we consider the design of the RJIO-SG and
RJIO-RLS algorithms equipped with the rank adaptation method
described in Section V.D. We consider $5$ interferers at $-60^{o}$,
$- 30^{o}$, $0^{o}$, $45^{o}$, $60^{o}$ with equal powers to the
SoI, which impinges on the array at $15^o$. Specifically, we
evaluate the rank adaptation algorithms against the use of fixed
ranks, namely, $D=3$ and $D=8$ for both SG and RLS algorithms. The
results show that the rank adaptation method is capable of ensuring
an excellent trade-off between convergence speed and steady-state
performance, as illustrated in Fig \ref{auto}. In particular, the
algorithm can achieve a significantly faster convergence performance
than the scheme with fixed rank $D=8$, whereas it attains the same
steady state performance.

\begin{figure}[!htb]
\begin{center}
\def\epsfsize#1#2{1\columnwidth}
\epsfbox{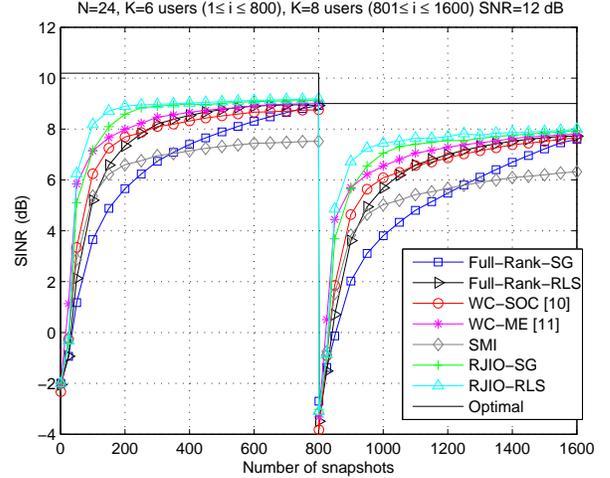} \vspace*{-1em} \caption{SINR performance of
robust LCMV algorithms against the number of snapshots with $M=24$,
$SNR=12$ dB in a non-stationary scenario} \label{ns}
\end{center}
\end{figure}

In the last experiment, we consider a non-stationary scenario where
the system has $6$ users with equal power and the environment
experiences a sudden change at time $i=800$. The $5$ interferers
impinge on the ULA at $-60^{o}$, $- 30^{o}$, $0^{o}$, $45^{o}$,
$60^{o}$ with equal powers to the SoI, which impinges on the array
at $15^o$. At time instant $i=800$ we have $3$ interferers with $5$
dB above the SoI's power level entering the system with DoAs
$-45^o$, $-15^o$ and $30^o$, whereas one interferer with DoA
$45^{o}$ and a power level equal to the SoI exits the system. The
RJIO and other analyzed \index{adaptive beamforming} algorithms are
equipped with rank adaptation techniques and have to adjust their
parameters in order to suppress the interferers. We optimize the
step sizes and the forgetting factors of all the algorithms in order
to ensure that they converge as fast as they can to the same value
of SINR. The results of this experiment are depicted in Fig.
\ref{ns}. The curves show that the RJIO algorithms have a superior
performance to the existing algorithms considered in this study.

\section{Conclusions}

We have investigated robust reduced-rank LCMV beamforming algorithms
based on robust joint iterative optimization of beamformers. The
RJIO reduced-rank scheme is based on a robust constrained joint
iterative optimization of beamformers according to the minimum
variance criterion. We derived robust LCMV expressions for the
design of the rank-reduction matrix and the reduced-rank beamformer
and developed SG and RLS adaptive algorithms for their efficient
implementation along with a rank adaptation technique. The numerical
results for an \index{adaptive beamforming} application with a ULA
have shown that the RJIO scheme and algorithms outperform in
convergence, steady state and tracking the existing robust full-rank
and reduced-rank algorithms at comparable complexity. The proposed
algorithms can be extended to other array geometries and
applications.

\newpage

\end{document}